\documentclass[prb,twocolumn,superscriptaddress,showpacs,amsmath,amssymb]{revtex4-2}
\usepackage[colorlinks=true, linkcolor=blue, filecolor=blue, urlcolor=blue, citecolor=blue]{hyperref}
\usepackage{amsfonts}
\usepackage{bbm}
\usepackage{graphicx}
\usepackage{dcolumn}
\usepackage{bm}
\usepackage{soul}

\begin{document}

\title{Four-Terminal Graphene-Superconductor Thermal Switch Controlled by The Superconducting Phase Difference}

\author{Peng-Yi Liu}
\affiliation{Department of Physics, Beijing Normal University, Beijing 100875, China}
\affiliation{International Center for Quantum Materials, School of Physics, Peking University, Beijing 100871, China}

\author{Yue Mao}
\affiliation{International Center for Quantum Materials, School of Physics, Peking University, Beijing 100871, China}

\author{Qing-Feng Sun}
\email[]{sunqf@pku.edu.cn}
\affiliation{International Center for Quantum Materials, School of Physics, Peking University, Beijing 100871, China}
\affiliation{Hefei National Laboratory, Hefei 230088, China}
\affiliation{CAS Center for Excellence in Topological Quantum Computation, University of Chinese Academy of Sciences, Beijing 100190, China}

\begin{abstract}
We propose a superconducting phase-controlled thermal switch
based on a four-terminal graphene-superconductor system.
By coupling two superconducting leads on a zigzag graphene nanoribbon,
both the normal transmission coefficient and crossed
Andreev reflection coefficient, which dominate the thermal conductivity of
electrons in the graphene nanoribbon, can be well controlled simultaneously
by the phase difference of the superconducting leads.
As a result, the thermal conductivity of electrons in the graphene nanoribbon
can be tuned and a thermal switching effect appears.
Using the non-equilibrium Green's function method, we verify
this thermal switching effect numerically. At ambient temperatures less than about one-tenth of the superconducting transition
temperature, the thermal switching ratio can exceed 2000.
The performance of the thermal switch can be regulated by the ambient temperature, and dopping or gating can slightly promote the thermal switching ratio. The use of narrower graphene nanoribbons and wider superconducting leads facilitates obtaining larger thermal switching ratios.
This switching effect of electronic thermal conductance in graphene is expected to be experimentally realized and applied.

\end{abstract}
\maketitle

\section{\label{sec1}Introduction}

A heat current can carry information and energy like an electric current.
Thermotronics \cite{thermotronics}, which corresponds to electronics,
is also expected to realize effects such as thermal diodes \cite{diode1,diode2,readd0}, negative differential thermal resistance \cite{negative1,negative2}, thermal transistors \cite{readd3}, thermal logic gates \cite{logic,readd4} and thermal memory \cite{storage1,storage2} based on heat currents.
However, compared to the control of electric currents, the ability to control the heat currents is much weaker. The electronics is well-developed, while the thermotronics has only started in the last two decades \cite{thermotrans}. In contrast to electronics, in thermal devices one does not have to worry about the deadly problem of heat dissipation.
In the design process of most thermal devices, it is crucial to control the on/off of the heat current, i.e., to make a thermal switch or thermal valve \cite{switch}.

A thermal switch can be defined as a device that connects a high temperature terminal (heat source) and a low temperature terminal (heat drain). By controlling some variables of the device, one can achieve significant control over the thermal conductivity of the device \cite{switchdef}. Many previous efforts have been made in the design of thermal switches, which can control thermal conductivity through many different variables, such as temperature itself \cite{temperature} and many kinds of phase transitions (solid-liquid, ferromagnetic-paramagnetic, semiconductor-metal, etc.) brought by temperature \cite{pt1,pt2,pt2s,pt3,pt4,pt5,pt6}, external electric or magnetic fields \cite{em1,em1d,em2,em3,em4,em5}, and the strain, pressure or twist of the materials \cite{sp1,sp2,sp3,sp4}, etc. The main carriers of heat current are also different, including phonons \cite{Phononics}, electrons \cite{thermotrans}, photons \cite{thermotrans,photonic} and even magnons \cite{magnetometers}, which makes the operating temperature of thermal devices cover an enormous range from milliKelvin to thousands of Kelvin, suitable for different occasions.

Superconductors are good thermal insulators at low temperatures \cite{refrigeration}, which naturally allows their application in the design of thermal devices.
The study of the superconducting-phase-controlled thermal transport, especially for electronic thermal transport, is known as phase-coherent caloritronics \cite{fast,pcc,readd0,readd3,readd4,TSQUIPT,theory}.
Very recently, by coupling a normal lead to a notch of a superconducting ring, Ligato {\sl et al.} experimentally demonstrated that the thermal conductivity of electrons of the normal lead can be significantly controlled by the magnetic flux of the superconducting ring \cite{TSQUIPT}.
Compared to other methods for thermal switch design, since the electronic temperature evolves much faster than the phononic temperature in superconducting systems \cite{refrigeration,fast}, phase-coherent caloritronics is expected to develop thermal switches that are more suitable for advanced applications, such as computing.

Graphene has gained much attention due to its unique band structure \cite{graphene0} in the last two decades.
Although the thermal transport of graphene is phonon-dominated over a large temperature range, with the decrease of temperature and device size, electronic thermal conductivity plays an increasingly important role \cite{graphenethermal1}. Due to the weak electron-phonon coupling, at low temperature, the heat in electrons is well isolated from phonons \cite{graphenethermal2,graphenethermal3,graphenethermal4}. So the temperature and thermal conductance of electrons in graphene can be measured and utilized almost unaffected by phonons. At the same time, graphene can be grown on an insulating substrate, ensuring that the heat of electrons is difficult to leak into the substrate. Therefore, it is possible to achieve excellent superconducting-phase-controlled electronic thermal transport	in a graphene-superconductor system.

In this paper, using a combination of graphene and superconductors, we propose
a thermal switch based on the idea of phase-coherent caloritronics.
The electric transport properties of the graphene-superconductor hybrid system
have been studied by a lot of literatures,
and many interesting phenomena have been exhibited,
such as the specular Andreev reflection \cite{sarl,four,addref1,addref2}
and the crossed Andreev reflection \cite{addref3}.
Here we study the thermal transport properties of electrons in a four-terminal graphene-superconductor system and propose a thermal switch device.
The thermal transport properties and influence factors of the system controlled by superconducting phase difference have been analyzed theoretically and calculated numerically.
By means of the non-equilibrium Green's function method,
it has been proved that the superconducting phase difference
can remarkably control the thermal conductance of electrons in the graphene nanoribbon, and the suitable working environments have been explored.
From our calculations, this four-terminal structure can achieve a thermal switching ratio of over 2000 under a low temperature at different on-site energy.
In addition, the production of graphene nanoribbons with widths below ten nanometers \cite{produce,produce2}, the high-quality contact between graphene nanoribbons and superconductors \cite{contact,contact2}, the control of the phase difference \cite{SQUIPT,TSQUIPT}, and the measurement of electronic temperature \cite{refrigeration,photonic,TSQUIPT,graphenethermal3} are all experimentally feasible, and thus our model is realizable. The theoretical studies and numerical calculations we have performed will advance the understanding of graphene-superconductor systems and the utilization of thermotronics.

The paper is organized as follows.
In Sec. \ref{sec2}, we present the model and theory studied in this paper, define the structure of the model and Hamiltonian, and derive the specific formulas of the thermal conductance by combining the non-equilibrium Green's function method with the Landauer-B\"{u}ttiker formula.
In Sec. \ref{sec3}, the numerical results of this paper are presented. Based on the numerical results, the control ability of the superconducting phase difference on the thermal conductance is discussed, and the effects of ambient temperature, on-site energy and the size of the system on the thermal switching ratio are studied.
Sec. \ref{sec4} concludes this paper.

\section{\label{sec2}Model and formulation}

\begin{figure}[!htb]
\centerline{\includegraphics[width=\columnwidth]{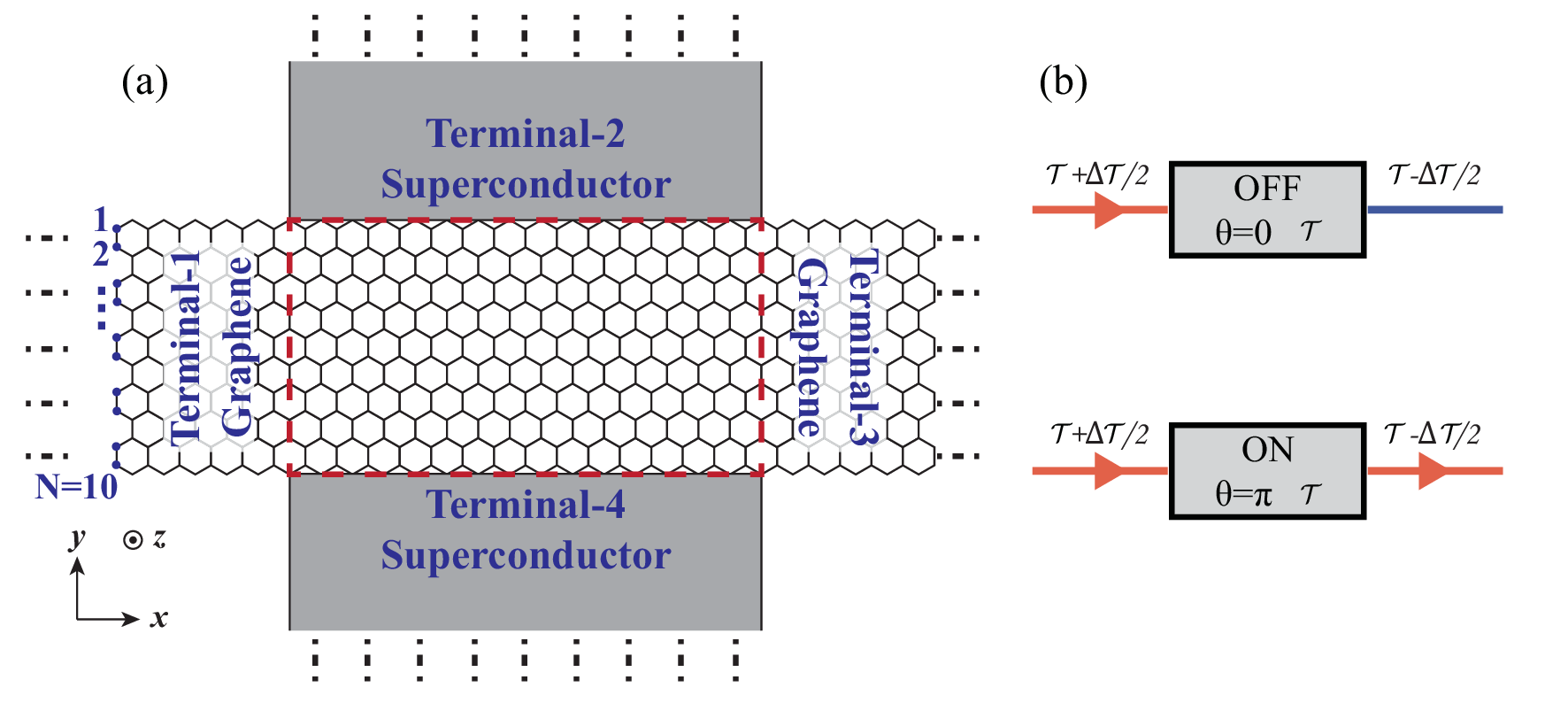}}
\caption{(a) Schematic diagram of the graphene-superconductor four-terminal thermal switch.
In this diagram, an infinite-length graphene nanoribbon symmetrically coupled two infinite-length superconducting leads. The width of the graphene nanoribbon is $N=10$ ($N$ is the number of zigzag chains, which are marked on the diagram) and the width of the superconducting leads is $W=31$ ($W$ is the number of atoms).
(b) Schematic diagram of working principle of thermal switch. When the superconducting phase difference $\theta=0$ ($\theta=\pi$), the heat current can hardly (easily) transmit from terminal-1 to terminal-3. The bias temperature between hot and cold terminal is $\Delta \mathcal{T}$, and the rest of system is at the ambient temperature $\mathcal{T}$.}\label{fig1}
\end{figure}

The thermal switching model we propose is a four-terminal graphene-superconductor system which consists of a zigzag graphene nanoribbon symmetrically coupled to two superconducting leads, as shown in Fig. \ref{fig1}(a).
We set that there exists a superconducting phase difference $\theta$ between two superconducting leads.
Experimentally, two superconducting leads can be connected to form a superconducting ring, so that the phase difference can be controlled by the magnetic flux through the ring \cite{SQUIPT,TSQUIPT}. The phase difference can also be tuned by a supercurrent flowing through two superconducting leads.
Fig. {\ref{fig1}}(b) shows the working principle of the thermal switch.
A bias temperature is set so that graphene terminal-1 has a high temperature initially. When the superconducting phase difference $\theta=0$, the thermal switch is switched off and the heat current can hardly flow from terminal-1 to terminal-3.
When $\theta=\pi$, the thermal switch is switched on and the thermal conductivity is significantly increased.

Using the nearest-neighbor tight binding method, the Hamiltonian of the system can be expressed as three parts \cite{four}. $H=H_G+\sum_{\alpha=2,4}(H_{S\alpha}+H_{T\alpha})$, where $H_G$, $H_{S\alpha}$ and $H_{T\alpha}$ are the Hamiltonians of
the graphene nanoribbon [including the dashed red box and the graphene leads of terminal-1/3 in Fig. {\ref{fig1}}(a)],
superconducting leads of terminal-$\alpha$, and the coupling between the superconducting lead terminal-$\alpha$ and the graphene nanoribbon, respectively, with $\alpha=2,4$. They can be expressed as \cite{four,addref4}
\begin{eqnarray}
H_{G}&=&\sum_{i\sigma}E_{0}a_{i\sigma}^{\dagger}a_{i\sigma}+\sum_{\left\langle ij\right\rangle\sigma}ta_{i\sigma}^{\dagger}a_{j\sigma},\label{HG}\\
H_{S\alpha}&=&\sum_{{\bm k}\sigma}\epsilon_{\bm k}b_{{\bm k}\sigma,\alpha}^{\dagger}b_{{\bm k}\sigma,\alpha}\nonumber\\
&+&\sum_{{\bm k}}\left(\Delta_{\alpha}^{*}b_{{\bm k}\uparrow,\alpha}^{\dagger}b_{-{\bm k}\downarrow,\alpha}^{\dagger}+\Delta_{\alpha}b_{-{\bm k}\downarrow,\alpha}b_{{\bm k}\uparrow,\alpha}\right),\label{HS}\\
H_{T\alpha}&=&\sum_{i\sigma}ta_{i\sigma}^{\dagger}b_{i\sigma,\alpha}+H.C.,\label{HT}
\end{eqnarray}
where $E_0$ and $t$ are the on-site energy and the hopping energy in the graphene region. $a_{i\sigma}^{\dagger}$ ($a_{i\sigma}$) and $b_{i\sigma,\alpha}^{\dagger}$ ($b_{i\sigma,\alpha}$) are the creation (annihilation) operators in the graphene and the superconductor with the spin $\sigma=\uparrow,\downarrow$ and the site index $i$.
Operators of superconducting leads can be transformed between momentum space and real space, $b_{i\sigma,\alpha}=\sum_{\bm k}b_{{\bm k}\sigma,\alpha}e^{i{\bm k}\cdot {\bm x}_i}$, where ${\bm x}_i$ is the coordinate at the site $i$.
We use $s$-wave superconductors for superconducting leads, with the superconducting gap $\Delta_\alpha=\Delta e^{i\phi_\alpha}$, and their phase difference $\theta=\phi_2-\phi_4$.

According to the Landauer-B\"{u}ttiker formula, the electric and heat current through terminal-1 can be obtained \cite{four,addref4,current,addref5}
\begin{eqnarray}
I_1&=&\frac{2e}{\hbar}\int \frac{{\rm{d}}E}{2\pi} [T_{12}\left(f_{1\rm e}-f_2\right)+T_{14}\left(f_{1\rm e}-f_4\right)\nonumber\\
&+&T_{13}^A\left(f_{1\rm e}-f_{3\rm h}\right)+T_{11}^A\left(f_{1\rm e}-f_{1\rm h}\right)\nonumber\\
&+&T_{13}\left(f_{1\rm e}-f_{3\rm e}\right) ],\label{I}\\
\dot{Q}_1&=&\frac{2}{\hbar}\int \frac{{\rm{d}}E}{2\pi} (E-\mu_1) [T_{12}\left(f_{1\rm e}-f_2\right)+T_{14}\left(f_{1\rm e}-f_4\right)\nonumber\\
&+&T_{13}^A\left(f_{1\rm e}-f_{3\rm h}\right)+T_{11}^A\left(f_{1\rm e}-f_{1\rm h}\right)\nonumber\\
&+&T_{13}\left(f_{1\rm e}-f_{3\rm e}\right) ],\label{Q}
\end{eqnarray}
where $e$ and $h$ represent the electron and the hole, respectively.
$\mu_1$ is the chemical potential of terminal-1.
We consider that a small bias voltage or a small bias temperature is applied
on the graphene nanoribbon between the terminal-1 and terminal-3 and
two superconductor terminals are set to be ground.
The Fermi distributions of the superconductor terminals are $f_{2(4)}=1/\left[ {\rm{exp}}(E/k_{B} \mathcal{T}_{2(4)})+1\right]$,
and the Fermi distributions in graphene terminals are
$f_{1(3)\rm e}=1/\{ {\rm{exp}}\left[ \left(E-eV_{1(3)}\right)/k_{B}\mathcal{T}_{1(3)}\right]+1\}$
and $f_{1(3)\rm h}=1/\{ {\rm{exp}}\left[ \left(E+eV_{1(3)}\right)/k_{B}\mathcal{T}_{1(3)}\right]+1\}$,
with the temperature $\mathcal{T}_{\alpha}$ and voltage $V_{\alpha}$.
When the bias voltage $\Delta V$ and the bias temperature $\Delta \mathcal{T}$ are infinitesimal, the Fermi distribution can be approximated linearly \cite{sim}.
Taking the electron as an example
\begin{eqnarray}
f_{\alpha \rm e}(E)=f_0(E)-eV_\alpha \frac{\partial f_0}{\partial E}+(\mathcal{T}_{\alpha}-\mathcal{T})\frac{\partial f_0}{\partial \mathcal{T}},\label{linear}
\end{eqnarray}
where $f_0=1/\left[ {\rm{exp}}\left(E/k_{B}\mathcal{T}\right)+1\right]$.
When we consider the electric conductance, we take $V_1=\Delta V$, $V_{2,3,4}=0$, $\mathcal{T}_{1,2,3,4}=\mathcal{T}$. When we consider the thermal conductance, we take $\mathcal{T}_{1,3}=\mathcal{T}\pm \Delta \mathcal{T}/2$, $V_{1,2,3,4}=0$, and $\mu_1=0$. We set all parts of the system except terminal-1/3 to maintain the ambient temperature $\mathcal{T}$.
Then, we can express the conductance as
\begin{eqnarray}
G=\frac{I_1}{\Delta V} &=&\frac{2e^2}{\hbar}\int \frac{{\rm{d}}E}{2\pi}\frac{1}{k_{B}\mathcal{T}}
 \frac{{\rm{exp}}(E/k_{B}\mathcal{T})}{[{\rm{exp}}(E/k_{B}\mathcal{T})+1]^2} \nonumber\\
&\times&(T_{12}+T_{14}+T_{13}+T_{13}^A+2T_{11}^A),\label{G}\\
\kappa=\frac{\dot{Q}_1}{\Delta \mathcal{T}}
&=&\frac{2}{\hbar}\int \frac{{\rm{d}}E}{2\pi} \frac{E^2}{k_{B}\mathcal{T}^2}\frac{{\rm{exp}}(E/k_{B}\mathcal{T})}
{[{\rm{exp}}(E/k_{B}\mathcal{T})+1]^2} \nonumber\\
&\times&(T_{12}/2+T_{14}/2+T_{13}+T_{13}^A).\label{kapa}
\end{eqnarray}
Different coefficients of transmissions and Andreev reflections in Eqs. (\ref{I}, \ref{Q}, \ref{G}, and \ref{kapa}) correspond to different physical processes, which can be calculated by the method of non-equilibrium Green's function \cite{four,addref4,current}.
\begin{itemize}
\item The normal transmission coefficient from terminal-1 to terminal-3, $T_{13}={\rm {Tr}}\left[{\mathbf {\Gamma}}_{1{\rm ee}}{\mathbf{G}}_{{\rm ee}}^r{\mathbf {\Gamma}}_{3{\rm ee}}{\mathbf {G}}_{{\rm ee}}^a\right]$;
\item The local Andreev reflection coefficient for the incident electron coming from terminal-1 with the hole Andreev reflected to terminal-1, $T_{11}^A={\rm {Tr}}\left[{\mathbf {\Gamma}}_{1{\rm ee}}{\mathbf {G}}_{{\rm eh}}^r{\mathbf {\Gamma}}_{1{\rm hh}}{\mathbf {G}}_{{\rm he}}^a\right]$;
\item The crossed Andreev reflection coefficient for the incident electron coming from terminal-1 with the hole Andreev reflected to terminal-3, $T_{13}^A={\rm {Tr}}\left[{\mathbf {\Gamma}}_{1{\rm ee}}{\mathbf {G}}_{{\rm eh}}^r{\mathbf {\Gamma}}_{3{\rm hh}}{\mathbf {G}}_{{\rm he}}^a\right]$;
\item The normal transmission coefficient from terminal-1 to a superconductor terminal, $T_{12(4)}={\rm {Tr}}\{{\mathbf {\Gamma}}_{1{\rm ee}}[{\mathbf {G}}^r{\mathbf {\Gamma}}_{2(4)}{\mathbf {G}}^a]_{{\rm ee}}\}$;
\end{itemize}
where ${\mathbf {\Gamma}}_\alpha$ is the matrix of the linewidth function
of the center region [the region surrounded by the dashed red box in Fig.{\ref{fig1}}(a)] coupled to the terminal-$\alpha$,
and ${\mathbf {G}}^{r(a)}$ is the matrix of the retarded (advanced) Green's function of the center region.
Here ${\mathbf {\Gamma}}_\alpha$ and ${\mathbf {G}}^{r(a)}$ have been expressed
in Nambu representation, and subscriptions $\rm{ee}$, $\rm{eh}$, $\rm{he}$, $\rm{hh}$ respectively represent the four matrix elements of Nambu subspace with $\rm{e}$ and $\rm{h}$ for the electron and hole components.

In order to calculate the coefficients above, we need the matrix of surface Green's function ${\mathbf{g}}_\alpha$ of each terminal, which can be obtained numerically \cite{surface}. Then, by combining the surface Green's functions with the coupling Hamiltonian between the center region and the terminal-$\alpha$, the retarded self-energy can be calculated, whose matrix elements can be expressed as $\Sigma_{ij,\alpha}^r=tg^r_{ij,\alpha}t$. And the linewidth function can be obtained together $\Gamma_\alpha=i[{\mathbf{\Sigma}}_\alpha^r-({\mathbf{\Sigma}}_\alpha^r)^{\dagger}]$. Finally, with the Dyson equation, the retarded and advanced Green's functions of the center region can be expressed as ${\mathbf{G}^r(E)}=[{\mathbf{G}^a}(E)]^{\dagger}=(E\mathbf{I}-\mathbf{H}_c-\sum_{\alpha=1,2,3,4}\mathbf{\Sigma}_\alpha^r)^{-1}$, where $\mathbf{H}_c$ is the Hamiltonian of the center region.

With the formulas in this section, the electric and thermal conductance can be calculated.
In the numerical calculations, we fix the superconducting gap $\Delta=1\rm{meV}$,
corresponding to a BCS transition temperature of about $T_c\approx \Delta/1.76k_B\approx6.6{\rm K}$, which is common for conventional superconducting transition temperatures, such as plumbum \cite{ssp}. The hopping energy of graphene is set to be $t=2.75\rm{eV}$ as a typical value \cite{graphene0}. The hopping energy $t_s=2.75{\rm eV}$ and lattice constants $a_s=\sqrt{3}\times 0.142{\rm nm}$ of the superconductors were selected to match the values of graphene, which lead to a similar BCS coherent length $\xi=\hbar v_{F}/\pi \Delta\approx500{\rm nm}$ to previous studies \cite{coherent} and much larger than the size of our system.	The Dynes broadening parameter \cite{addDynes} (i.e. the imaginary part added to the energy when calculating the Green's function) is set to $\gamma=10^{-6}\Delta$, and it can be verified that the result calculated by this value is almost the same as those calculated with the common value $\gamma=10^{-4}\Delta$ \cite{TSQUIPT,theory}.

\section{\label{sec3}Results and Discussions}

In this section, the numerical calculations of the above model are performed,
and the control of the superconducting phase difference on the thermal conductance of the system is given. In addition, the effect of ambient temperature $\mathcal{T}$, the on-site energy $E_0$ (dopping or gating) and the size of the center region $(N,W)$ on the performance of the thermal switch is investigated.
\begin{figure}[!htb]

\centerline{\includegraphics[width=\columnwidth]{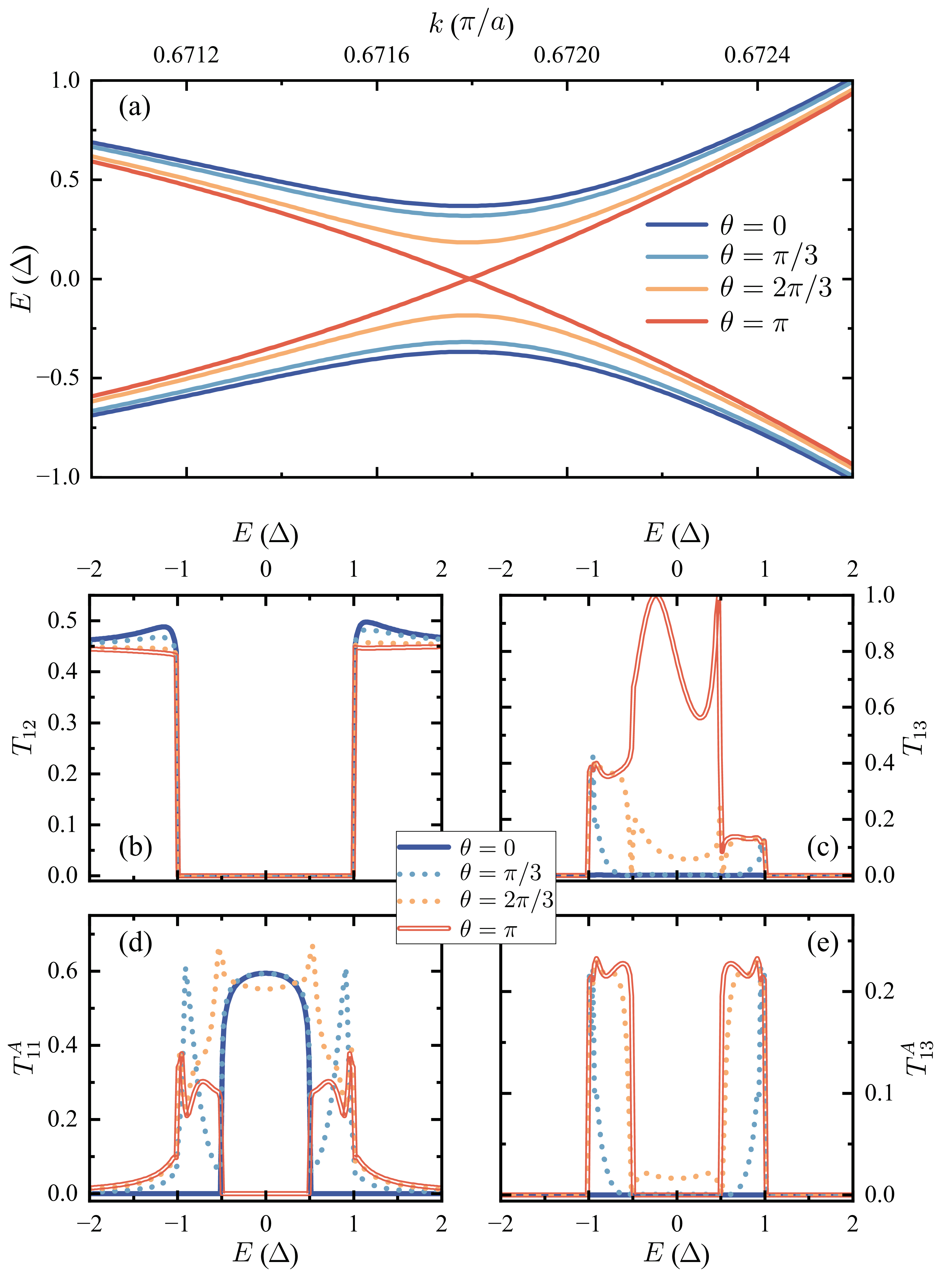}}
\caption{The influence of superconducting phase difference $\theta$ on the energy gap and several coefficients. Here the phase difference $\theta$ is set to be $0$, $\pi/3$, $2\pi/3$ and $\pi$ respectively.
(a) The band structure near the energy gap of the proximitized graphene. We infinitely extend the proximitized center region and two superconductors (discrete square lattices with the width of 28 lattices) in the horizontal ($\pm {\bm x}$) direction to obtain the band structure. (b), (c), (d), and (e) are the normal transmission coefficient $T_{12}(E)$, the normal transmission coefficient $T_{13}(E)$, the local Andreev reflection coefficient $T_{11}^A (E)$, and the crossed Andreev reflection coefficient $T_{13}^A (E)$ versus the incident energy $E$, respectively. Parameters involved in the calculation are taken as: the width of graphene nanoribbon $N=40$, the width of superconducting leads $W=201$, and the on-site energy $E_0=-0.5\Delta$.}
\label{fig2}
\end{figure}

In Eqs. (\ref{G}) and (\ref{kapa}), we can see that the control of superconducting phase difference $\theta$ on the electric and thermal conductance is equivalent to the control of these transmission and Andreev reflection coefficients.
Therefore, we need to investigate the effect of the superconducting phase difference on each coefficient, and the results are shown in Fig. {\ref{fig2}}.
First, in Fig. {\ref{fig2}(a)}, we present the band structure near the gap of the center graphene region [the dashed red box in Fig. {\ref{fig1}(a)}] proximitized by superconductors, which dominates the thermal switching effect.
When calculating the energy band, we do not consider terminal-1 and terminal-3, and extend the center region and two superconductors coupled on the upper and lower sides (a SNS junction) infinitely in the horizontal ($\pm {\bm x}$) direction. Here, we use the upper and lower superconductors with discrete square lattices and the 28 lattices in the $y$ direction to provide the proximity effect. Thus, the horizontal wave vector $k$ becomes a good quantum number to obtain the solution of the energy band.
The results show that, due to the superconducting proximity effect,
the band opens a gap near the Fermi surface controlled by the superconducting phase difference $\theta$. When $\theta=0$, the gap is the largest, while the gap is closed when $\theta=\pi$, which is consistent with the results of previous studies on SNS junctions \cite{sns}. Electrons and holes from terminal-1 need to tunnel through a barrier to transmit to terminal-3, when the gap is opened.

Fig. {\ref{fig2}}(b) shows the normal transmission coefficient from graphene terminal-1 to superconducting terminal-2 $T_{12} (E)$. Obviously, $T_{12}$ appears only outside the superconducting gap $\Delta$. The image of $T_{14}(E)$ (not given) is similar to $T_{12}(E)$, and due to the geometric symmetry of our system, $T_{12} (\theta)=T_{14} (-\theta)$ exactly.
It can be seen that since the normal transmission into the superconductor involves only one superconducting terminal, the effect of the phase difference $\theta$ on $T_{12}$ is not significant.
The normal transmission coefficient from graphene terminal-1 to graphene terminal-3 $T_{13} (E)$ is shown in Fig. {\ref{fig2}}(c). As the superconducting phase difference increases in $[0,\pi]$, the gap of the system gradually closes, and the barrier height that electrons need to tunnel through decreases \cite{TSQUIPT, sns}. Therefore, $T_{13}$ increases dramatically with increasing $\theta$. $T_{13}\approx 0$ when $\theta=0$, while $T_{13}$ is close to $1$ when $\theta=\pi$, which achieves a good switching effect.

Figs. {\ref{fig2}}(d) and (e) show the coefficients of the local Andreev reflection $T_{11}^A (E)$ and the crossed Andreev reflection $T_{13}^A (E)$, respectively.
Since our system has the joint symmetry combined by the time-reversal operator with the mirror reflection about the $xz$ plane
or with the rotation of $\pi$ about the $z$ axis \cite{sym1,sym2},
the Andreev reflection coefficients $T_{11}^A$ and $T_{13}^A$
are symmetrical about $E=0$.
With the increase of the phase difference $\theta$ in $[0,\pi]$,
$T_{11}^A$ is reduced to $0$ when the incident energy $\lvert E \rvert \le \lvert E_0 \rvert$, and $T_{13}^A$ increases from 0 for $\lvert E_0 \rvert \le \lvert E \rvert \le \Delta$,
as a result of the constructive and destructive interference of reflected holes.
This is because the superconductor-graphene interface can occur both intraband ($\lvert E \rvert \le \lvert E_0 \rvert$) Andreev retro-reflection (ARR) and interband ($\lvert E_0 \rvert \le \lvert E \rvert \le \Delta$) specular Andreev reflection (SAR) \cite{sarl,sarr}.
When $\theta=\pi$, ARR is eliminated by destructive interference and only SAR occurs. In contrast to ARR, the amplitude of SAR in a zigzag graphene nanoribbon with even chains has odd-parity under mirror operation \cite{four,parity}.
As a result, when $\theta=0$, SAR is eliminated by destructive interference,
and ARR reaches the maximum value.
Due to the quantum diffraction effects, the reflected holes do not have definite directions, which makes holes of SAR (ARR) can also transmit to terminal-1(3), leading to the nonzero value of $T_{11}^A$ ($T_{13}^A$) in the region of SAR (ARR).
However, considering the size of the center region,
the barrier is long and multiple reflections can happen.
Therefore, $T_{13}^A$ within $\lvert E \rvert \le \lvert E_0\rvert$ is always much smaller than $1$, and increases almost monotonically with
the phase difference $\theta$ in $[0,\pi]$ [see Figs. {\ref{fig2}}(e)].
In addition, $T_{13}$, $T_{13}^A$ have no tail outside $\Delta$, while $T_{11}^A$ has, which is because the first two are lost by multiple reflections, during the process of reaching terminal-3.

To sum up the results of Fig. {\ref{fig2}}, the superconducting phase difference $\theta$ significantly controls $T_{13}$, $T_{11}^A$, and $T_{13}^A$,
which mainly occur within the superconducting gap $\Delta$.
Through the control of the gap at the center region and the interference of reflected holes, $T_{13}$ and $T_{13}^A$ increase almost monotonically as $\theta$ increases in $[0,\pi]$. But the effect of phase difference on $T_{12}$ and $T_{14}$ is weak, which occurs only outside the superconducting gap. Next, combined with Eqs. (\ref{G}) and (\ref{kapa}), we can calculate and analyze the effect of phase difference
on electric and thermal conductance.

\begin{figure}[!htb]
\centerline{\includegraphics[width=\columnwidth]{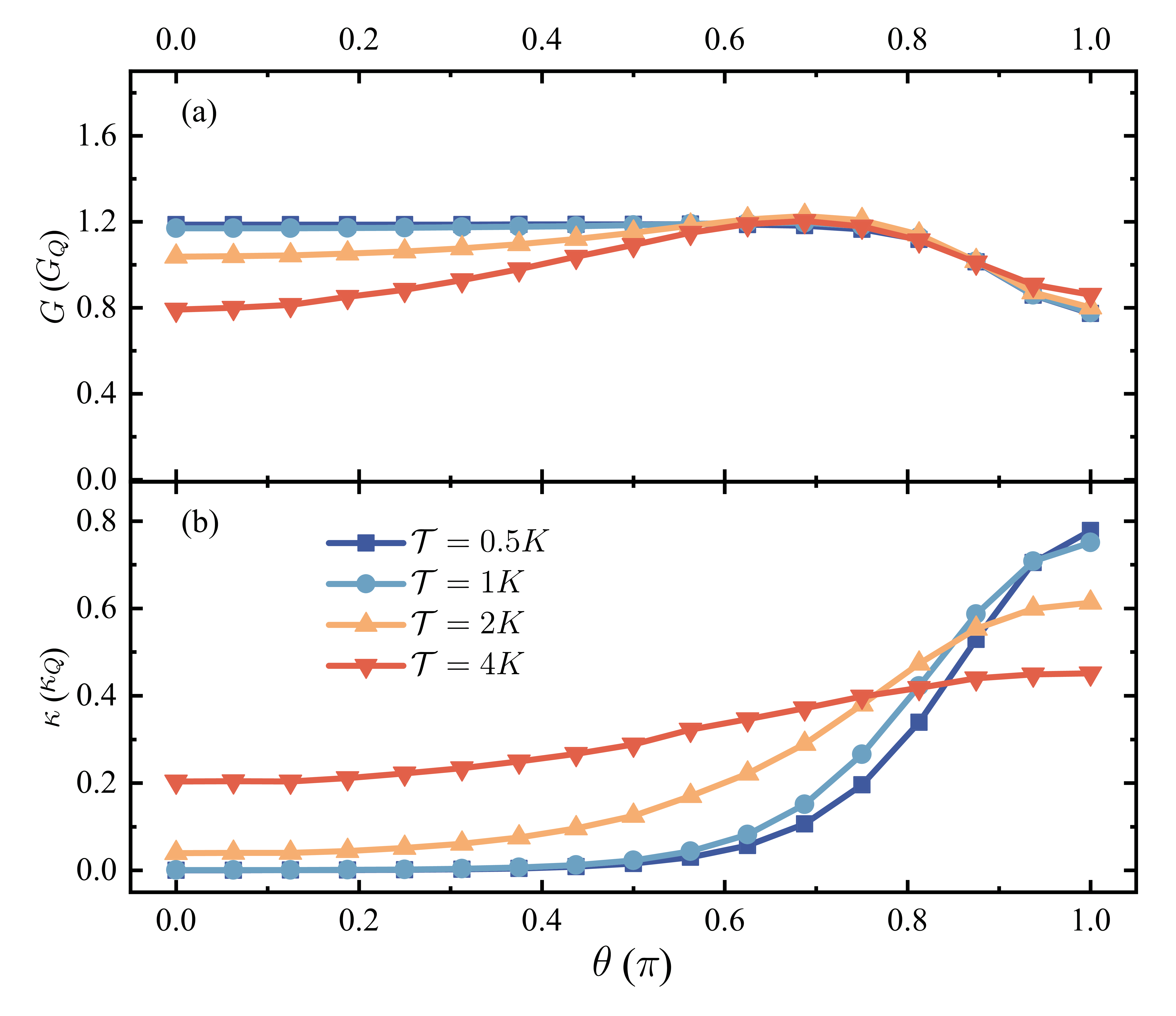}}
\caption{(a) Electric conductance $G$ and (b) thermal conductance $\kappa$
versus superconducting phase difference $\theta$ for four different ambient temperatures $\mathcal{T}=0.5{\rm K}$, $1{\rm K}$, $2{\rm K}$ and $4{\rm K}$.
The calculated results are unitized by quantum electric conductance $G_Q=2e^2/h$ and quantum thermal conductance $\kappa_Q=2\pi^2 k_B^2 \mathcal{T}/3h$.
Parameters involved in the calculation are taken as: the width of graphene nanoribbon $N=40$, the width of superconducting leads $W=201$, and the on-site energy $E_0=-0.5\Delta$.}
\label{fig3}
\end{figure}

In Fig. {\ref{fig3}}, we show the variation of electric conductance $G(\theta)$
and thermal conductance of electrons $\kappa(\theta)$ with the superconducting
phase difference $\theta$.
It can be seen that the control of phase difference on each coefficient does
not induce a good electric switch effect.
This is because the superconducting phase difference $\theta$
has opposite control effects on $T_{11}^A$ and $T_{13}, T_{13}^A$.
As $\theta$ increases from 0 to $\pi$, $T_{13}, T_{13}^A$ increase while $T_{11}^A$ decreases [shown in Fig. {\ref{fig2}(c-e)}].
As a result, we cannot observe the phase-controlled effect on the conductance $G$
because that $G$ is proportional to the sum of $2 T_{11}^A$ and $T_{13}, T_{13}^A$
as shown in Eq. (\ref{G}).
From the physical picture, this is because Cooper pairs are able to conduct electric currents in superconductors without being restricted by the energy gap of the center region.
As opposed to the conductance $G$, the superconducting phase difference
achieves a significant on/off control of the thermal conductance at the appropriate temperature, similar to recent experiments \cite{TSQUIPT}.
This is because as $\theta$ increases from $0$ to $\pi$, both the normal transmission coefficient $T_{13}$ and the crossed Andreev reflection coefficient $T_{13}^A$ are enhanced:
Because of the energy gap closing, $T_{13}$ significantly increases from almost zero; Due to the constructive interference of reflected holes, $T_{13}^A$ also increases.
From the physical picture, the Cooper pairs cannot conduct heat \cite{refrigeration,thermotrans},
so that the thermal conductance $\kappa$ is independent of
the local Andreev reflection coefficient $T_{11}^A$ [see Eq. (\ref{kapa})]
and $\kappa$ can be well controlled by the energy gap of the center region.
In addition, it can be seen in Fig. {\ref{fig3}}(b) that ambient temperature has a great influence on the performance of the thermal switch, which can be measured by the thermal switching ratio $r=\kappa_{max}/\kappa_{min}$ \cite{switchdef}.
Next, we study the influence of ambient temperature and
other variables on thermal switch performance by calculating $r$.

\begin{figure*}[!htb]
\centerline{\includegraphics[width=2\columnwidth]{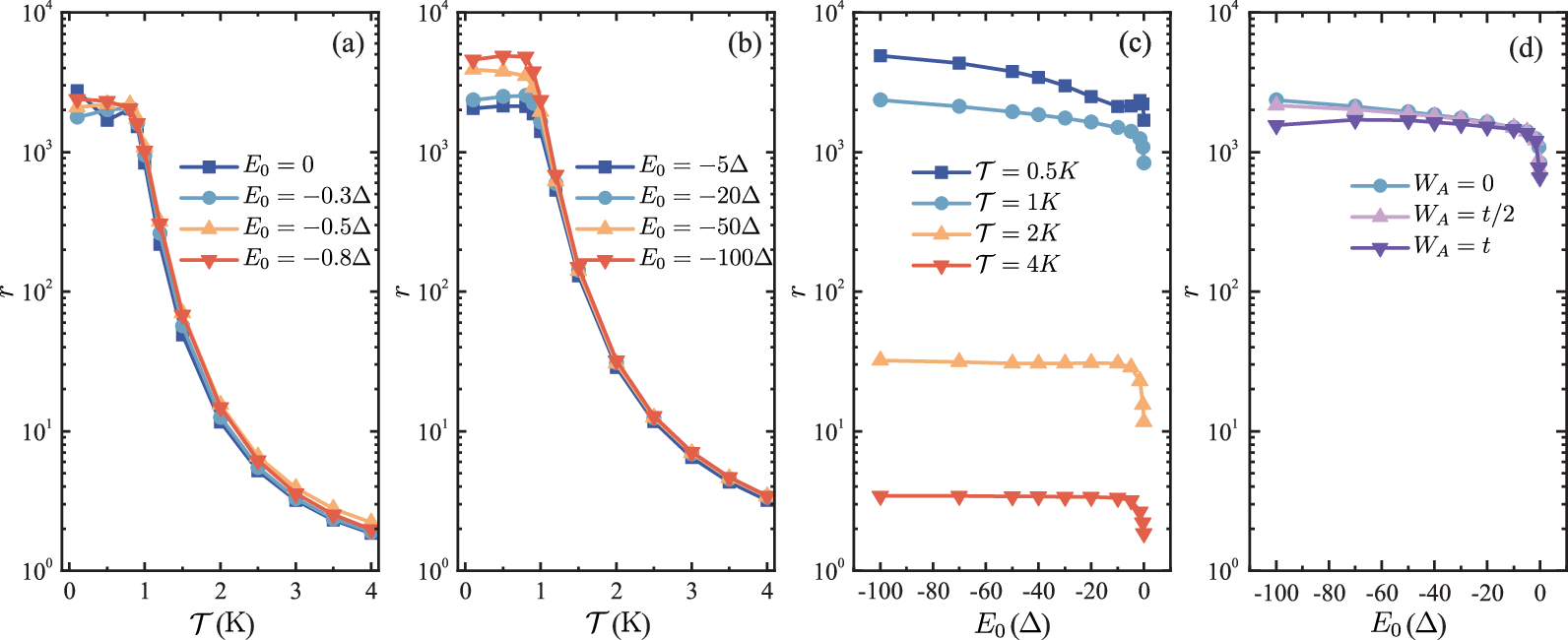}}
\caption{
(a) and (b) The influence of ambient temperature $\mathcal{T}$
on thermal switching ratio $r$ with the different on-site energy $E_0$.
(c) The influence of on-site energy $E_0$ on thermal switching ratio $r$
at four different ambient temperatures $\mathcal{T}$.
(d) Thermal switching ratio $r$ versus the on-site energy $E_0$
in the presence of the Anderson disorder with three different disorder strength $W_A$. The temperature is fixed at $\mathcal{T}=1{\rm K}$, and each point is averaged by 200 configurations.
In the calculation of (a-d), we use the width of graphene nanoribbon $N=40$
and the width of the superconducting leads $W=201$.
}
\label{fig4}
\end{figure*}

The calculation results of thermal switching ratio varying
with ambient temperature $r(\mathcal{T})$ is shown 
in Figs. {\ref{fig4}}(a) and {\ref{fig4}}(b).
The thermal switch we devise mainly controls the heat current flowing in the graphene nanoribbon, which depends on the good thermal insulating performance of the superconductor at low temperatures \cite{TSQUIPT}.
As the temperature rises, the Fermi distribution widens.
As a result, more electrons with energy outside $\Delta$ are involved.
We already know from the above discussion that the electrons outside $\Delta$ are mainly transmitted normally into the superconductors and are hardly controlled by the phase difference, as shown in Fig. {\ref{fig2}}(b).
Therefore, as shown in Figs. {\ref{fig4}}(a) and {\ref{fig4}}(b),
with the increase of ambient temperature, the thermal switching ratio drops sharply with all on-site energies, which is consistent with the results of recent experiments \cite{TSQUIPT}.
When $\mathcal{T}\leq 1{\rm K}$ (about $k_B \mathcal{T} \leq \Delta/10$),
the thermal switch maintains good performance.
For example, the thermal switching ratio $r$ can exceed 2000
at $\mathcal{T} =0.5{\rm K}$, which is much larger than that given by previous thermal switches (their ratio is usually less than 100) \cite{pt1,pt2s,pt6,em1d}. Our considerations here are based on conventional superconductors ($\Delta=1{\rm {meV}}$), but this device is also suitable for high-$T_c$ superconductors.
With the large energy gap of high-$T_c$ superconductors,
the sensitivity to temperature will be significantly reduced.

On the contrary, the increase of $\left|E_0\right|$
not only does not harm the thermal switching effect,
but also slightly promotes its performance,
and this part of the calculation results are shown in Fig. {\ref{fig4}}(c).
For either the normal transmission or the crossed Andreev reflection,
the incident electron or hole from the terminal-1 needs to pass through the barrier in the center region caused by the gap in order to reach terminal-3.
Since the change in the on-site energy will not have a sufficient effect on the amplitude of the gap, our thermal switch can work
at a large range of the on-site energies $E_0$.
In particular, the thermal switching ratio $r$ can always exceed 2000
regardless of $E_0$ at the low temperature
[see the curve with $\mathcal{T}=0.5{\rm K}$ in Fig.\ref{fig4}(c)].
Increasing $\left|E_0\right|$ can increase the density of states of graphene,
which is experimentally reflected in doping or adjusting the gate voltage,
to increase the carrier concentration
and then to enhance electronic thermal conductivity in the on and off states.
However, the increase of the carrier concentration also
enhances the superconducting proximity effect on graphene,
so that the control of thermal conductivity by superconductors increases as well.
As shown in Fig.\ref{fig4}(c), with the increase of $\left|E_0\right|$, the thermal switching ratio initially (when $\left|E_0\right|$ is small) has a rapid increase and soon tends to saturation.

In addition to the case of ballistic transport,
we can further consider the presence of the impurities
and diffusive transport caused by impurities and doping.
We consider the case of Anderson disorder by adding a uniform random
distributed term $w_i\in \left[-W_A/2,W_A/2\right]$ to the on-site energy
in the center graphene region, i.e. $H_{center}=\sum_{i\sigma}(E_{0}+w_i)a_{i\sigma}^{\dagger}a_{i\sigma}
+\sum_{\left\langle ij\right\rangle\sigma}ta_{i\sigma}^{\dagger}a_{j\sigma}$,
to simulate the effect of disorder on the system \cite{addref1}.
As shown in Fig.\ref{fig4}(d), we fix the temperature $\mathcal{T}=1{\rm K}$, calculate and redraw the corresponding curve in Fig.\ref{fig4}(c).
With the increase of the disorder strength $W_A$, the thermal switching ratio
is slightly reduced, not surprisingly.
However, the large thermal switching ratios survive even under a strong
disorder for $W_A=t=2.75{\rm eV}$, which shows that the proposed thermal switch
is not sensitive to disorder.

\begin{figure}[!htb]
\centerline{\includegraphics[width=\columnwidth]{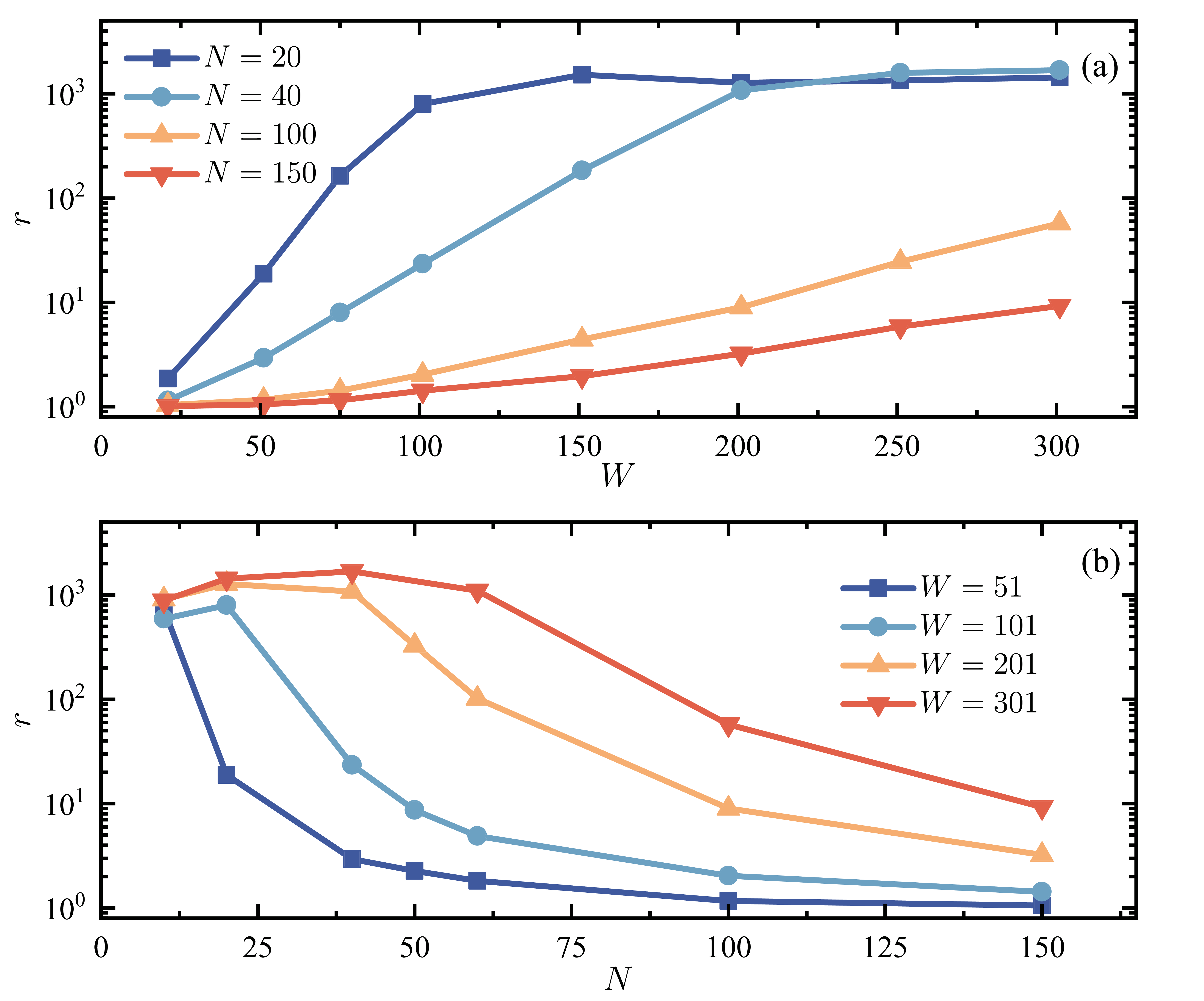}}
\caption{The influence of the system size on thermal switching ratio $r$.
(a) The influence of the width $W$ of superconducting leads on thermal switching ratio $r$ at four different $N$.
(b) The effect of graphene width $N$ on thermal switching ratio $r$
at four different $W$. Parameters involved in the calculations are the ambient temperature $\mathcal{T}=1{\rm K}$ and the on-site energy $E_0=-0.5\Delta$.}
\label{fig5}
\end{figure}

Last, let us study the influence of the size of the system
on the performance of the thermal switch.
In the discussion of Fig. {\ref{fig2}}, we mention that
when the phase difference $\theta=0$,
the opened energy gap becomes a barrier through which electrons and holes
need to pass in the center region.
The size of the system has noticeable effects on the barrier.
The height of the barrier, i.e., the amplitude of the gap opening
in the center region, decreases exponentially with increasing the width of the graphene nanoribbon $N$. And the barrier width is approximately equal to the width of the superconducting leads $W$.
Therefore, decreasing $N$ and increasing $W$ will inhibit heat current from passing through the center region
when the energy gap is open at $\theta=0$,
which will allow the thermal switch to close better.
On the contrary, when $\theta=\pi$, the gap closes and the barrier vanishes.
In the absence of an energy gap at the center region,
changing $N$ and $W$ will have almost no effect on the opening of the thermal switch. Therefore, the thermal switching ratio will decrease with increasing $N$ and increase with increasing $W$. Fig. {\ref{fig5}} shows the results of the calculation for varying the system size. The variation of the thermal switching ratio is highly consistent with our analysis. In general, the choice of narrower graphene nanoribbons and wider superconducting leads allows the thermal switch to work better.

\section{\label{sec4}Conclusions}

In summary, we propose a superconducting phase-controlled
thermal switch which consists of a graphene nanoribbon coupled
by two superconducting leads.
When the superconducting phase difference decreases from $\pi$ to $0$,
the normal transmission coefficient
(crossed Andreev reflection coefficient) decreases significantly to 0
due to the opening of the energy gap (destructive interference),
then the thermal conductance of electrons in graphene nanoribbon also decreases significantly.
Using the non-equilibrium Green's function method,
we have verified this thermal switching effect numerically.
The results show that, the thermal switching ratio can exceed 2000,
at ambient temperatures less than about one-tenth of the superconducting transition temperature.
The performance of the thermal switch can be regulated
by the ambient temperature, while 
the increase of the carrier concentration can slightly
promote the thermal switching ratio.
A narrower graphene nanoribbon and wider superconducting leads are favorable
for obtaining a larger thermal switching ratio.
In addition, the production of graphene nanoribbons
with width below ten nanometers,
the high-quality contact between graphene nanoribbons and superconductors,
the control of the superconducting phase difference and the measurement of electronic temperature
are all experimentally feasible,
which provides conditions for realizing our model and high thermal switching ratio.
Our work deepens the understanding of graphene-superconductor systems
and provides theoretical support for the research of thermotronics.

\section*{\label{sec5}ACKNOWLEDGMENTS}

This work was financially supported by National Natural Science Foundation of China (Grant No. 12374034 and No. 11921005), the Innovation Program for Quantum Science and Technology (2021ZD0302403), and the Strategic Priority Research Program of Chinese Academy of Sciences (Grant No. XDB28000000).


\end{document}